\documentclass[fleqn,usenatbib]{mnras}

\usepackage{newtxtext,newtxmath}

\usepackage[T1]{fontenc}

\DeclareRobustCommand{\VAN}[3]{#2}
\let\VANthebibliography\thebibliography
\def\thebibliography{\DeclareRobustCommand{\VAN}[3]{##3}\VANthebibliography}

\usepackage{graphicx}	
\usepackage{amsmath}	

\usepackage{bm}
\usepackage{subfigure}
\usepackage{hyperref}

\newcommand{\dd}{ {\textrm d}}
\newcommand{\ee}{ {\textrm e}}

\title[Application of KK Theory in Compact Stars]{Application of Kaluza--Klein Theory in Modeling Compact Stars: Exploring Extra Dimensions}

\author[A. Horv\'ath, E. Forgács-Dajka, G.G. Barnaföldi]{
Anna Horv\'ath,$^{1,2}$\thanks{E-mail: horvath.anna@wigner.hun-ren.hu (AH)}
Emese Forgács-Dajka,$^{2,3}$
and Gergely G\'abor Barnaf\"oldi$^{1}$
\\
$^{1}$HUN-REN Wigner Research Center for Physics, 29-33 Konkoly-Thege Mikl\'os Str., H-1121 Budapest, Hungary\\
$^{2}$ELTE E\"otv\"os Lor\'and University, Institute of Physics and Astronomy, Department of Astronomy, H-1117 Budapest, P\'azm\'any P\'eter s\'et\'any 1/A, Hungary\\
$^{3}$HUN-REN-SZTE Stellar Astrophysics Research Group, H-6500 Baja, Szegedi út, Kt. 766, Hungary
}

\date{Accepted XXX. Received YYY; in original form ZZZ}

\pubyear{\the\year{}}

\begin{document}
\label{firstpage}
\pagerange{\pageref{firstpage}--\pageref{lastpage}}
\maketitle

\begin{abstract}

A theoretical framework for calculating the mass-radius curve of compact stars in the Kaluza\,--\,Klein space-time is introduced, with one additional compact spatial dimension. Static, spherically symmetric solutions are considered, with the equation of state provided by a zero temperature, interacting multidimensional Fermi gas. To model the strong force between baryons, a repulsive potential is introduced, which is linear in the particle number density. The maximal mass of compact stars is calculated for different model parameters, and with a physical parameter choice, it satisfies observational data, meaning that it is possible to model simple, realistic objects within this framework. Based on this comparison, a limiting size for the observational regime of extra dimensions in compact stars is provided, with $r_c \gtrsim 0.2$~fm. 

\end{abstract}





\section{Introduction} 
\label{sec:intro}

Depending on the mass of the progenitor star, compact objects become white dwarfs, neutron stars or black holes after gravitational collapse. Intermediate mass stars evolve into neutron stars through type II supernova explosions; these make up magnetars, fast-rotating pulsars -- the densest objects of our Universe still exhibiting matter properties. The name "compact" indicates their high mass-to-size ratio: they have a typical radius of about a dozen kilometers, while they weigh a few solar masses. These exotic objects can be studied via multi-messenger astrophysical observations, thus providing an excellent playground for testing novel theoretical models at the most extreme physical regimes, which are unavailable to state-of-the-art technologies of mankind. The nomenclature of neutron stars is pretty rich: on one hand, the family members are named after their special observational properties, such as fast-rotating pulsars, or magnetars with high magnetic fields. On the other hand, categories arise from differences in theoretical modeling, reflecting the composition of the exotic matter interior: strange stars, hyperon stars, quark stars, quark-hybrid stars, gravastars, dark-matter stars. Similarly to these, Kaluza\,--\,Klein compact stars have been introduced by~\cite{Kan:2002yi,Lukacs:2003fh,Barnafoldi:2007}. 
It is natural to model these objects not only within the standard model of particle physics, but open the gates to novel theoretical approaches, since their extraordinary parameters -- small size, high mass, stable fast rotation, high energy and gravitational radiation -- suggest extreme physics. What supports this challenging task is the surprisingly high precision of the main compact star observables: mass, rotational period and tidal deformability (see examples below, in Table~\ref{tab:maxmass}). The size of these celestial bodies also plays a key role, however, measuring this property is rather model dependent, requiring {\sl a priori} assumptions on the surface emission and the matter of the star~\cite{Demorest_2010,Fonseca_2021}.    

Several efforts in the context of spacetime with extra dimensions have already been made. The original idea leads back to the 1920s, when Th. Kaluza~\cite{Kaluza:1921tu} and O. Klein~\cite{Klein:1926tv} (KK) unified gravity and electromagnetism in a unified five dimensional theory by introducing an extra compactified spatial dimension. Later, in 1998, N. Arkani-Hamed, S. Dimopoulos and G. Dvali~\cite{ARKANIHAMED1998263} (ADD) developed the theory further by the assumption that only gravitons are allowed to travel in the extra dimension, letting gravity leak out, making it less strong. A year later, in 1999, L. Randall and R. Sundrum~\cite{PhysRevLett.83.3370} (RSI) predicted the geometry of space beyond our brane to be warped, affecting our measurements of gravity's strength by making it appear weaker.~\footnote{We note, that many other theories contributed to the development of this field, including non-extradimensional models, such as~\cite{PhysRev.124.925} as well.} These ideas of (large) compactified extra dimensions would result in several benefits in high-energy physics by: explaining the matter-to-antimatter ratio, resolving the hierarchy problem, and explaining the phenomenon of dark matter. As of today, no clear experimental evidence has been found to support the existence of the theoretically predicted extra dimensions, however, constraints and limitations can be obtained by combining high-energy experiments, cosmological observations and astrophysical data, as listed below:
\begin{description}
    \item[Grand Unified (GUT) and Brane theory:] The aim of many modern theories of physics is to unify the fundamental interactions and to solve the hierarchy problem. We know that eletromagnetism and the weak force merge at the electroweak scale, however, grand unified theories predict the unification of all non-gravitational interactions. Gravity is much weaker, than the quantum mechanical forces (hierarchy problem), and has a different, geometrical description. It is expected to become comparable in strength with the interactions described by the standard model (SM) of particle physics at the Planck scale, $l_\mathrm{P}=\sqrt{\hbar G /c^3} = 1.616 \cdot 10^{-35}$~m. The hierarchy problem can also be resolved by theories, where standard model particles exist on a brane embedded in extra-dimensional space, in which gravity dilutes. In these models, the size of the compact dimensions is less than $10^{-17}$~m~\cite{Sundrum_1999,PhysRevLett.83.3370}.
    \item[TeV-scale and Large Extra Dimensions (LED)] The energy scale at which gravity might be unified with other forces is reduced down to the TeV-scale by the presence of (large) extra dimensions in the ADD concept. Today's particle accelerators have reached this energy regime with high-luminosity, therefore they enable us to test if extra dimensions exist. To obtain this result, one method is to estimate the production of “microscopic black holes”, which would depend on the number of extra dimensions, and determine their mass, size and dimensions -- indeed, the energy at which they form. If microscopical black holes do appear in the reaction created in hadron-hadron collisions, they evaporate rapidly within about $10^{-27}$~s, by decaying into standard model or supersymmetric particles. Such events should contain an exceptionally large multiplicity (number of tracks) or signs for new particles, which would not appear without the presence of extra dimensions. Collisions always create energy-balanced events, but if e.g. a graviton would be created and able to propagate into extra dimensions, that would cause imbalance with missing energy in the detectors. Searches like these have been ongoing for decades, and as the integrated luminosity is increasing at colliders, constraints are getting more accurate. The Large Electron-Positron (LEP) collider provided $r_c <1.9\cdot 10^{-4}$~m for the case of two extra dimensions~\cite{Cheung_2002,Bernardi2003LowSG,DELPHI:2008uka}. Fermilab has improved these values based on data from the D0 and CDF experiments at the Tevatron, whose figures ($r_c <3.3\cdot 10^{-4}$~m) were consistent with the measurements of DESY's HERA electron-proton collider at the H1 experiment~\cite{Adloff_2000}. Another method is related to the so-called Kaluza\,--\,Klein states. If known particles could propagate in extra dimensions, their heavier versions would hold the same properties (e.g. quantum number) as their Standard Model counterparts, thus could be measured in the usual way, however, with a larger mass. CMS and ATLAS experiments at the Large Hadron Collider (CERN) are searching for e.g. 100-times more massive $Z$- or $W$-like bosons in TeV-scale proton-proton collisions as signatures of extra dimensions. Current experimental data places upper limits on the size of extra dimensions in the sub-millimeter range~\cite{Giudice:2000av,Nath:2010zj,Khachatryan_2011,Aaboud_2016,Beuria:2017jez,ATLAS_2024}.
    \item[Precision measurements of gravity] Newton’s law at short distances might require corrections ("the fifth force"), which can be approximated by a Yukawa-like effective term, whose parameters may be determined by the properties of the curled extra dimensions~\cite{Fischbach:1986ka}. Performing torsion pendulum experiments and atomic interferometry can put constraints on the size of extra dimensions. Using this method, the University of Washington's Eöt-Wash group has excluded the existence of extra dimensions larger than $8.0\cdot 10^{-5}$~m, regardless of how many dimensions there are~\cite{Adelberger:2009zz}. Similar long-term experiments are ongoing with an enhanced-precision Eötvös-pendulum at the Jánossy Underground Research Laboratory at HUN-REN Wigner Research Centre for Physics~\cite{Eotvos:1922pb,Szemle_2022}. 
    A more precise measurement using gravitational wave interferometers by placing a dynamic gravitational field generator (DFG) at their axis was proposed by~\cite{Raffai_2011,Famaey_2012,Murata:2014nra}.
    \item[Astrophysical and cosmological observations] The existence of extra dimensions might have observable effects on the behavior of gravity at large scales, therefore modifying the dynamics of the universe, or perturbing the formation, propagation and polarization of gravitational waves~\cite{LIGOScientific:2021sio}. 
    The stability of the universe, fluctuations of the cosmic microwave background (CMB) radiation, the gravitational wave background (GWB), and structure formation in cosmology provide indirect constraints on the existence of extra dimensions. As an example: gravity is much stronger near a black hole, than it would be without extra dimensions, so the brightness of stars would appear to be larger, than it otherwise would, which can be presented with 44\% increase for star S2 in Sagittarius A at the center of the Milky Way~\cite{Bin_Nun_2010,Bin_Nun_2011}. Another study demonstrated that the shapes of extra dimensions can influence the cosmic energy released by the violent birth of the universe 13.8 billion years ago. Current limits based on gravitational wave observations suggest that the sizes of compactified extra dimensions are smaller than the $10^{-6}$~m scale~\cite{Carroll_2001,Shiu_2007,Andriot_2017,Yu_2019}.

\end{description}

\begin{table*}
\centering
\caption{Close to maximal mass neutron star observations with high accuracy.}
\begin{tabular}{lcccr}
\hline
Constellation & Name & Mass, [$M_\odot$] & Detection device & Reference \\ 
\hline
\hline
    Scorpius & PSR\,J1614$-$2230 & $1.908^{+0.0016}_{-0.0016}$  & Green Bank & \cite{Arzoumanian} \\
    Scorpius & PSR\,J1614$-$2230 & $1.97^{+0.04}_{-0.04}$  & Green Bank & \cite{Demorest_2010} \\
    Taurus & PSR\,J0348$+$0432  & $2.01^{+0.04}_{-0.04}$  & Arecibo, Effelsberg, Green Bank & \cite{Antoniadis:2013pzd} \\
    Camelopardalis & PSR\,J0740$+$6620 &  $2.08^{+0.07}_{-0.07}$ & CHIME, Green Bank &     \cite{Fonseca_2021} \\  
    Hercules & PSR\,J1810$+$1744 & $2.13^{+0.04}_{-0.04}$ & Keck I  & \cite{Romani_2021} \\
    Lacerta & PSR\,J2215$+$5135 & $ 2.27^{+0.17}_{-0.15} $ &  IAC-80, William Herschel  & \cite{Linares_2018}  \\ 
    Columba & PSR\,J0514$-$4002E & $2.35^{+0.20}_{-0.18}$ & MeerKAT  & \cite{Barr:2024wwl} \\ 
    Sextans &  PSR\,J0952$-$0607 & $2.35^{+0.17}_{-0.17}$  & Keck I    & \cite{Romani_2022} \\
    Sculptor & GW190814 & $2.59^{+0.08}_{-0.09}$  & LIGO/VIRGO  & \cite{Abbott_2020} \\
\hline
\end{tabular}

\label{tab:maxmass}
\end{table*}

Overall, existing experimental data suggest that if extra dimensions are real, their effects must be confined to scales significantly smaller than those typically associated with the usual 4-dimensional spacetime. Consequently, the size of extra compactified dimensions is generally constrained to be below $10^{-6}$~m.

According to our knowledge as of today, besides high-energy particle accelerators and cosmological observations, maximal mass neutron stars can also be good candidates for extra dimension searches. The more relativistic a star is, the higher the chance it enhances relativistic effects: spacetime curvature, compactness and interplay between fundamental interactions towards grand unified theories. A massive neutron star can have highly energetic degrees of freedoms in its core, where one can explore the interplay between the equilibrium state degenerated extreme matter and the gravitational force indirectly. The list of the highest mass objects ($M\gtrsim 1.9-2.0~M_{\odot}$), which can serve as potential candidates for our challenge are collected in Table~\ref{tab:maxmass}. We also included special observations, where the pulsar or low-mass black hole status of the object is still under discussion~\cite{Abbott_2020,Barr:2024wwl}.

We think, such compact star observations may further tighten the bounds, therefore in this work, we aim to provide constraints and physical parameter values for an extra dimensional theory. We use the simplest interacting multidimensional Fermi gas model for the first time in Kaluza\,--\,Klein spacetime with one microscopical compactified extra dimension. We set the interaction strength according to nuclear forces in a typical maximal mass neutron star, and present a realistic Kaluza\,--\,Klein compact star. Comparison of the calculated macroscopic stellar parameters with observational data may restrict the properties of extra dimensions within this model framework. 

The paper is organized as follows. We briefly introduce the theoretical background of compact stars modelled in Klauza\,--\,Klein spacetime and the equation of state (EoS) in Section~\ref{sec:KK}. The calculations of the EoS is described in Section~\ref{sec:eos}, along with some plots with respect to the three main parameters of the model of the interacting multi-dimensional Fermi gas. Obtaining the mass-radius relation is required to perform the solution of the Tolmann\,--\,Oppenheimer\,--\,Volkoff equations, which are given in Section~\ref{sec:tov}. In Section~\ref{sec:disc}, we describe the results and the consequences of the microscopic parameters of the interacting Fermi gas and the extra compactified dimensions' size for the macroscopic stellar observables. Finally, we conclude with a summary in Section~\ref{sec:sum}.


\section{Compact stars in the Kaluza\,--\,Klein space-time -- theoretical background}
\label{sec:KK}

In order to arrive at observable quantities on an astronomical scale, one needs to connect the microscopical, nuclear physics properties of matter to the structure of compact objects. The Kaluza\,--\,Klein model, that we are investigating, affects physics on the nuclear scale, thus it introduces modifications to the equations describing matter, and through a statistical physical treatment it appears in the equation of state. Astrophysical objects in the general relativistic regime can be modeled by the Tolman\,--\,Oppenheimer\,--\,Volkoff (TOV) equation, which describes hydrostatic equilibrium with relativistic corrections in a static, spherically symmetric spacetime. The TOV equation gives the change in pressure with respect to the radius of the star at each point using the local value of the energy density, and that is how thermodynamics (EoS) is introduced in the structure. In the current section we describe the building blocks of such a procedure, then combine them and show how macroscopic properties of compact objects are calculated.

\subsection{The Kaluza\,--\,Klein space-time and geometry}

\begin{figure}
    \centering
    \includegraphics[width=0.9\linewidth]{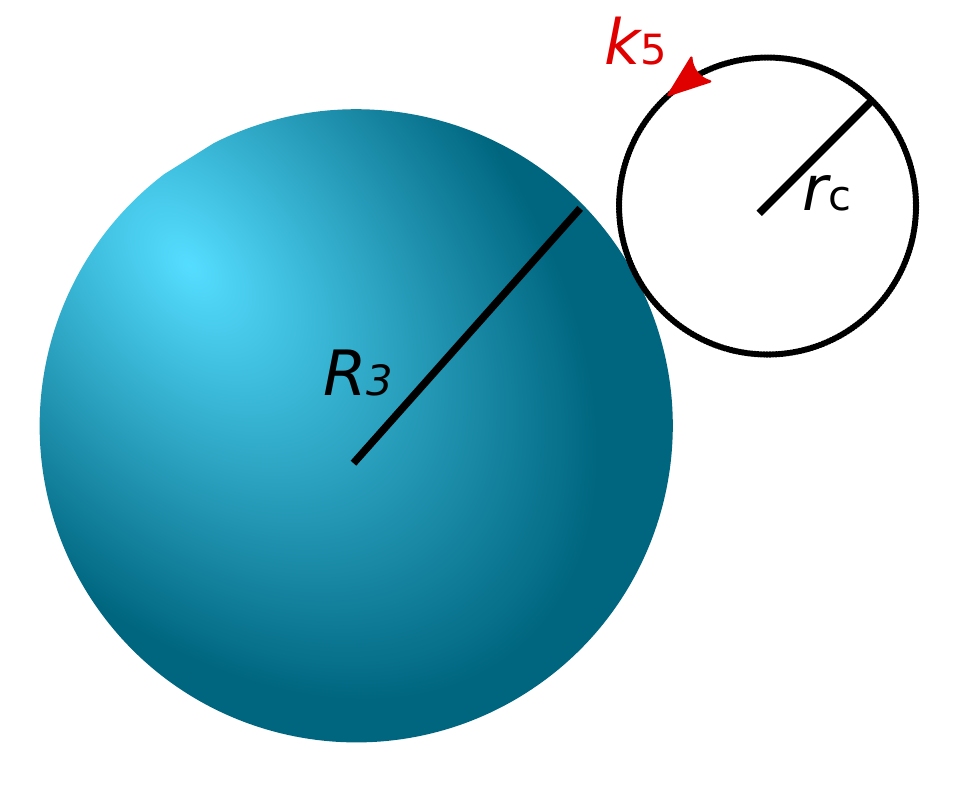}
    \caption{Schematical view of a projection of the Kaluza\,--\,Klein spacetime structure.}
    \label{fig:kk_sphere}
\end{figure}

The Kaluza\,--\,Klein theory is dated back to the beginning of the 20\textsuperscript{th} century, when\,---\,inspired by general relativity\,--\,Theodor Kaluza proposed a geometrical description to electromagnetism, unifying it with gravity in a five dimensional metric tensor in \cite{Kaluza:1921tu}. One of the key elements of this solution was to introduce the so-called {\it cylinder condition}, which means that the components of the metric do not depend on the 5\textsuperscript{th} coordinate. Later, Oskar Klein suggested that the 5\textsuperscript{th} dimension should be microscopic and compactified in~\cite{Klein:1926tv}. This gives a straightforward quantum-mechanical-motivated explanation to why it does not have an effect at the large-scales. By this interpretation spacetime is made of the usual 1 timelike and 3 infinite, large-scale spatial dimensions, with the addition of one spatial compactified microscopic extra dimension, curled up in a circle, originating at each point in usual space (see Fig.~\ref{fig:kk_sphere}).    

The general form of the Kaluza\,--\,Klein metric with the applied signature ($+----$) can be written as 
\begin{equation}
    g_{AB} =\left[\begin{matrix}g_{\alpha\beta}+\kappa^2\Phi^2A_\alpha A_\beta & \kappa\Phi^2 A_\alpha \\\kappa\Phi^2 A_\beta & \Phi^2 \end{matrix}\right]\ ,
    \label{eq:general_metric}
\end{equation}
where $g_{AB}$ is the five and $g_{\alpha \beta}$ is the four dimensional metric. Capital Roman letters run over 0, 1, 2, 3 and 5, while Greek ones run from 0 to 3, as usual\footnote{Note, traditionally the 4\textsuperscript{th} spatial coordinate is denoted by index 5 for being of the fifth compactified dimension.}. $A_\alpha$ can be associated with the four-vector containing the vector and scalar potentials of electromagnetism, while $\Phi$ represents a scalar field and $\kappa$ is an arbitrary constant following \cite{Overduin:1997sri}. One can consider a simple model, where the effects of magnetic fields are not taken into account, and we also neglect the electric field generated by charge polarization, since it does not affect the maximal mass as in \cite{inproceedings}, which is in our main focus. 
Thus we neglect $A_\alpha$ altogether. Since our focus is on compact astrophysical objects (neutron stars, pulsars, etc.) in hydrostatic equilibrium, spacetime is assumed to be static and spherically symmetric, thus the five dimensional metric can be written in a diagonalized form
\begin{equation}
    g_{AB}=\left[\begin{matrix}g_{00} & 0 & 0 & 0 & 0\\0 & g_{11} & 0 & 0 & 0\\0 & 0 & g_{22} & 0 & 0\\0 & 0 & 0 & g_{22} \sin^{2}\theta  & 0\\0 & 0 & 0 & 0 & g_{55}\end{matrix}\right]\ , 
    \label{eq:diagmetric}
\end{equation}
where $\theta$ is the polar angle. 

\begin{figure}
    \centering
    \includegraphics[width=0.95\linewidth]{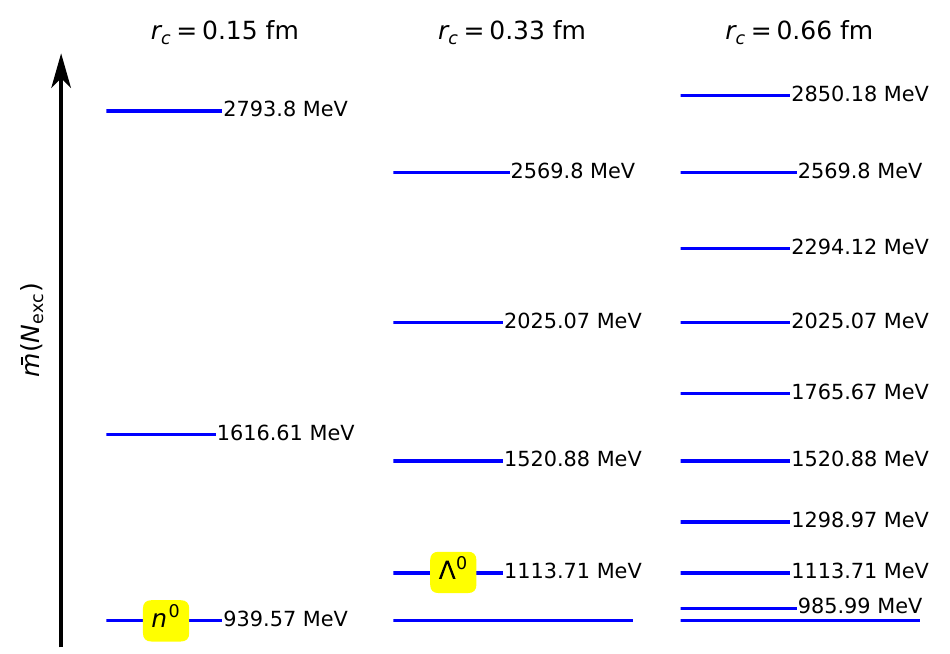}
    \caption{Kaluza\,--\,Klein ladder: excited states of the neutron ($n^0$) shown for different $r_c$ values. Calculating with $r_c=0.33$~fm, the first excited state corresponds to the $\Lambda^0$ baryon.}
    \label{fig:kk_ladder}
\end{figure}

By considering the 5\textsuperscript{th} dimension, a new degree of freedom is introduced, which can be interpreted in both a geometrical and a field theoretical, rather mathematical way~\cite{delCastillo:2020wka}. We assume particles are allowed to propagate in large extra dimensions, which size (compactification radius, $r_c$) is at the order of hundreds of fm. 
 Since the size of the fifth dimension is finite, particles moving in it obey a periodic boundary condition, which leads to discrete, quantized momentum spectrum called the \textit{Kaluza\,--\,Klein ladder}. 
The existence of these infinite number of excitations in the Kaluza\,--\,Klein theory is often considered problematic~\cite{Lacquaniti:2009yy}, however, here we use it to model the hadronic mass spectrum of particles (see Fig.~\ref{fig:kk_ladder}) with respect to the Particle Data Group~\cite{PhysRevD.110.030001}. 

The energy of a particle moving in five dimensional spacetime using natural units ($c=1=\hbar$) can be written as
\begin{equation}
    E = \sqrt{\mathbf{k}^2+k_\mathrm{5}^2+m^2} = \sqrt{\mathbf{k}^2+\left(\frac{N_\mathrm{exc}}{r_\mathrm{c}}\right)^2+m^2}\ ,
    \label{eq:energy}
\end{equation}
where $m$ is the rest mass of the particle in 5D, $\mathbf{k}$ is the standard, 3-dimensional momentum, $N_\mathrm{exc}$ is the excitation number (the energy level filled on the KK ladder) and $r_\mathrm{c}$ is the size, the radius of the extra compactified spatial dimension. Thus the fifth dimensional component of the momentum is
\begin{equation}
    k_\mathrm{5} = \frac{N_\mathrm{exc}}{r_\mathrm{c}}\ .
\end{equation}
Each momentum component plays a similar role in Eq.~\eqref{eq:energy} as the mass, thus one can interpret $k_\mathrm{5}$ as a contribution to the particle's mass in the usual 1+3-dimensional picture
\begin{equation}
    \bar{m}^2(N_\mathrm{exc}) = m^2 + k_\mathrm{5}^2\ ,
\end{equation}
indeed leading to a 1+3D effective mass, $\bar{m}$.

Figure~\ref{fig:kk_ladder} presents the Kaluza\,--\,Klein ladder for different compactification radius sizes ($r_\mathrm{c}=$ 0.15, 0.33 and 0.66~fm). We have chosen the neutron mass as the ground state of the spectra, thus $m=m_{n^0}=939.57$~MeV, and the vertical axis represents excited states with mass, $\bar{m}$. The smaller the radius $r_c$, the larger the gap between each energy level, and, on the contrary, excited states approach continuum for large $r_\mathrm{c}$ values. Thus, it is manifest that particles can propagate in the extra dimension only if they possess a sufficient amount of energy. This is the reason why in normal, everyday conditions the effects of the fifth dimension are negligible, however, in more extreme circumstances, such as in the interior of compact stars~\cite{Nattila:2022evn} or in the early Universe, significant deviations from usual physics can be predicted. For example, we expect the appearance of more massive hadronic states, which can be associated with strangeness flavor\footnote{Kaluza\,--\,Klein theory gives rise to a charge, $Q \propto N_\mathrm{exc}\cdot \sqrt{G}/r_\mathrm{c}$, which can be associated with the symmetry of moving in the extra compactified dimension with a periodic boundary condition. The usual interpretation of this charge is the electric one, however, one can also think of it as strangeness as in~\cite{Gegenberg,Fischbach:1986ka,Lukacs:2003fh}.} to obtain the next neutral baryon state 
$\Lambda^0$, which is denoted on the Kaluza\,--\,Klein ladder in Figure~\ref{fig:kk_ladder}. One can see that with a proper choice of the compactification radius $r_c$, it can be produced as the first excitation of the neutron. We note, further excited states can also be associated with more-and-more massive hadronic states, such as $\Lambda_c$, $\Lambda_b$, etc. -- as pointed out in~\cite{Lukacs:2003fh,Barnafoldi:2007,Barnafoldi:2015wca}. Taking into account observations in general, in our study we assume, that the size ($r_c$) of large extra dimensions may play role in a compact star is at the order of 0.01-100~fm.

Kaluza\,--\,Klein theory is the first, simplest example of a physical model considering extra dimensions. Many other theories of modern physics~\cite{Overduin:1997sri}, such as certain models of modified gravity, eg. scalar-tensor theories~(\cite{PhysRev.124.925}), $f(R)$ gravity~\cite{10.1093/mnras/150.1.1, STAROBINSKY198099}, or quantum gravity models, eg. string theory~\cite{Veneziano:1968yb}, M-theory~\cite{WITTEN199585} are based on, or can be connected to Kaluza\,--\,Klein in effective ways, depending also on the interpretation. However, with many of these theories, it is difficult to make predictions, and test if the existence of extra dimensions is possible. Here we propose a simple method with which qualitative effects of extra dimensions may be tested using neutron star measurements. 

\subsection{Fermi gas in multidinemsional space with repulsive interaction}

The Kaluza\,--\,Klein theory can be connected to the macroscopic observables of a compact object through its effects on thermodynamics (EoS). We chose to model the matter inside a neutron star using the equation of state of the Fermi gas. However, for the case of the extra dimensional Kaluza\,--\,Klein picture, we have to generalize the model in more than three spatial dimensions. The thermodynamic potential in $d$ spatial dimensions is
\begin{eqnarray}
    \label{eq:termpot}
    \Omega =&&-V_{(d)}\sum\limits_{i=0}^{N_\mathrm{exc}} \frac{g_i}{\beta}\int\limits \frac{\dd^d \textbf{k}}{(2\pi)^d} \times \nonumber \\
     &&\times 
    \left[\ln \left(1+\ee^{-\beta(E_i-\mu)}\right)+\ln \left(1+\ee^{-\beta(E_i+\mu)}\right) \right]\ ,
    \end{eqnarray}
where $V_{(d)}$ is the $d$-dimensional volume, $i$ runs over particles with different excitation numbers ($N_\mathrm{exc}$), $g_i$ is the multiplicity, $\beta=1/k_BT$, $\mathbf{k}$ is the $d$-dimensional momentum vector, $E_i$ and $\mu$ are the energy and the chemical potential respectively. After this point, we set $k_B=1$ as well. The two exponential terms are associated with particles and antiparticles, respectively. For neutron stars, the zero temperature limit is a good approximation, since, due to the high density, all fermionic states are filled, see~\cite{Glendenning:1997wn}. In the zero temperature limit antiparticle states are not present and one can arrive at the approximation
\begin{equation}
T\ln \left. \left(1+\ee^{-(E-\mu)/T}\right) \right|_{T=0} = \left \{ 
\begin{array}{cl}
\mu-E, & \textrm{if}\;\;  E<\mu  \\ 
0, &  \textrm{if}\;\;E\geq \mu\ . 
\end{array}
 \right. \nonumber
\end{equation} 
From the 3-dimensional thermodynamic potential the state variables of the non-interacting degenerate Fermi gas with $g=2$ can be derived:
\begin{equation}
    \begin{aligned}
        \varepsilon_0&=\frac{1}{4\pi^2}\left[\mu k_\mathrm{F}\left(\mu^2-\frac{1}{2}m^2 \right)-\frac{1}{2}m^4 \log\left(\frac{\mu+k_\mathrm{F}}{m} \right) \right]\ , \\
        p_0&=\frac{1}{12\pi^2}\left[\mu k_\mathrm{F}\left(\mu^2-\frac{5}{2}m^2 \right)+\frac{3}{2}m^4 \log\left(\frac{\mu+k_\mathrm{F}}{m} \right) \right]\ , \\
        n_0&=\frac{k_\mathrm{F}^3}{3\pi^2}\ , 
    \end{aligned}
\end{equation}
where $\varepsilon_0$ is the energy density, $p_0$ is the pressure, $n_0$ is the baryon number density and $\mu = (m^2+k_\mathrm{F}^2)^{1/2}$, where $k_F=(\mu^2-m^2)^{1/2}$ is the Fermi momentum. The chemical potential at zero temperature in equilibrium is equal to the Fermi energy. 
Note, the mass of particles for a Fermi gas in the Kaluza\,--\,Klein model can be associated with the mass of the given excited-states, $\bar{m}(N_\mathrm{exc})$, when measured in the 3-dimensional space. 

So far, we have considered only the usual description of a non-interacting Fermi gas. However, at nuclear densities and above, the contribution of the strong force between particles cannot be neglected, indeed, it would lead to neutron stars whose maximal mass cannot reach the measured maximum values~\cite{Lukacs:2003fh,Karsai:2016wfx}. 

Thus we use a simple potential 
\begin{equation}
    U(n) = \xi n \ ,
    \label{eq:poti}
\end{equation}
that is linear in the number density, $n$ to model the strong force in its repulsive regime~\cite{Zimanyi:1987bt}, where $\xi$ is a constant that could be fixed by matching it to mean-field models at lower densities, however, in this article we leave it to be a parameter that can be set in a phenomenological way. The potential will contribute to the pressure and the energy density with a correction term
\begin{equation}
    p_{int}=\varepsilon_{int}=\int U(n)dn = \frac{1}{2}\xi n^2
\end{equation}
and to the chemical potential as $\bar{\mu}=\mu-U(n)$ such that
\begin{equation}
    \begin{aligned}
        p(\mu)&=p_0(\bar\mu)+p_{\mathrm{int}} \\
        \varepsilon(\mu)&=\varepsilon_0(\bar\mu)+\varepsilon_{\mathrm{int}}\ .
        \label{eq:ep}
    \end{aligned}
\end{equation}
The number density needs to be calculated using a numerical root finding method, since
\begin{equation}
        n(\bar{\mu}) = n_0(\bar{\mu})=\frac{k_\mathrm{F}(\bar{\mu})^3}{3\pi^2}=\frac{1}{3\pi^2}\left[ \left( \mu-U(n) \right)^2-m^2 \right]^{3/2}\ .
\end{equation}

Considering the model from the 1+3 dimensional perspective, we have the hadronic spectrum of particles. Thus, we simply take the thermodynamic potential Eq.~\eqref{eq:termpot} and sum over the spectrum created by the presence of the extra dimension. The chemical potential will be equal for all particles. The first step is to calculate the total number density, regardless of excitation number, from which we have the modified chemical potential $\bar{\mu}$. Then, we can calculate the partial pressure and energy density using Eqs.~\eqref{eq:ep} for each excitation and then take their sum to get the corresponding total pressure and energy density, which can be used in the Tolman\,--\,Oppenheimer\,--\,Volkoff equation.  

We note, that additional effects of the extra dimension could be taken into account. The appearance of this extra degree of freedom could modify the phase space, thus the equation of state 
in further, non-trivial ways~\cite{Wojnar:2022dvo}, including a possible modification to Heisenberg's uncertainty principle~\cite{Petruzziello:2021vyf}. This is also connected to the overall question, whether the strong gravitational field of a neutron star can be neglected at a microscopic level, or not, thus affecting the equation of state~\cite{1995IJTP...34.1843K}.
The definition of movement in the compact dimension, thus the physical interpretation of the 5-dimensional geodesic equation~\cite{Lacquaniti:2009yy} and the dispersion relation are not trivial either. A further effect could come from a more complex interaction between particles propagating in the extra dimension~\cite{Lukacs:2003fh}. In this study our focus is on investigating the most evident consequences of the existence of a fifth dimension, so we neglect these additional effects. 


\section{Results on the Kaluza\,--\,Klein compacts star} 
\label{sec:result}

Our aim here is to investigate the properties of the derived equation of state (EoS) compared to some other selected models of compact star interior e.g. from~\cite{compose2}. Then, we discuss how one can compute the properties of compact stars, taking into account possible complications. Finally, we compare results to measured, astrophysical observational data as well. 

\subsection{Calculating the equation of state of a Kaluza\,--\,Klein star}
\label{sec:eos}

\begin{figure}
    \centering
    \includegraphics[width=0.95\linewidth]{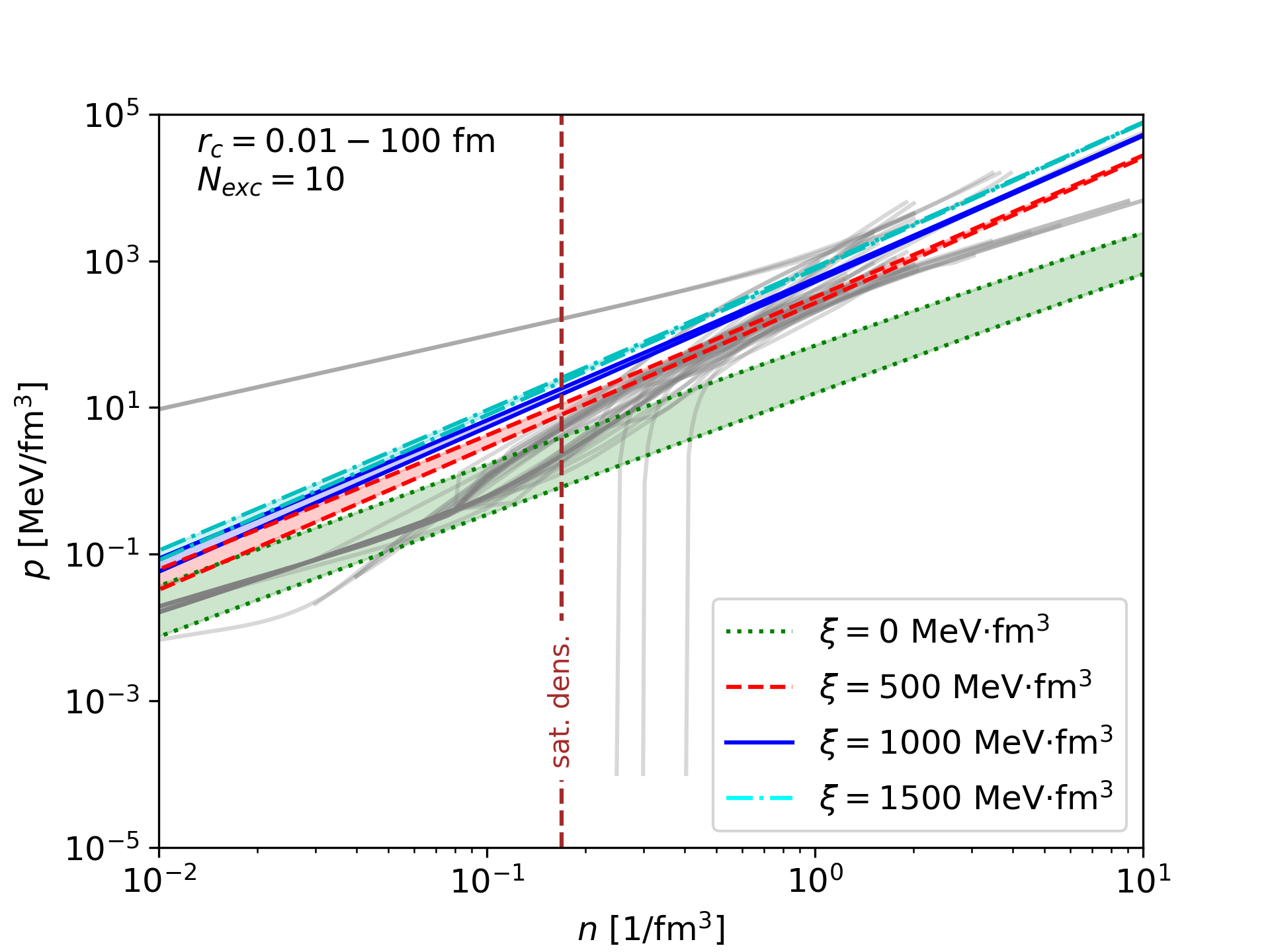}
    \caption{Equation of state: pressure as a function of total baryon number density on a logarithmic scale. Colors and line styles correspond to different $\xi$ interaction strengths. Bands span $r_c$ values between 0.01~fm (lower limit) and 100~fm (upper limit). The nuclear saturation density is shown (dashed brown vertical line), as well as EoSs taken from other works for comparison (light grey curves)~\citep{Ozel:2016oaf,compose2}.
    }
    \label{fig:sat}
\end{figure}

\begin{figure*}
     \centering
     \hfill
     \begin{subfigure}
        \centering
         \includegraphics[width=0.45\linewidth]{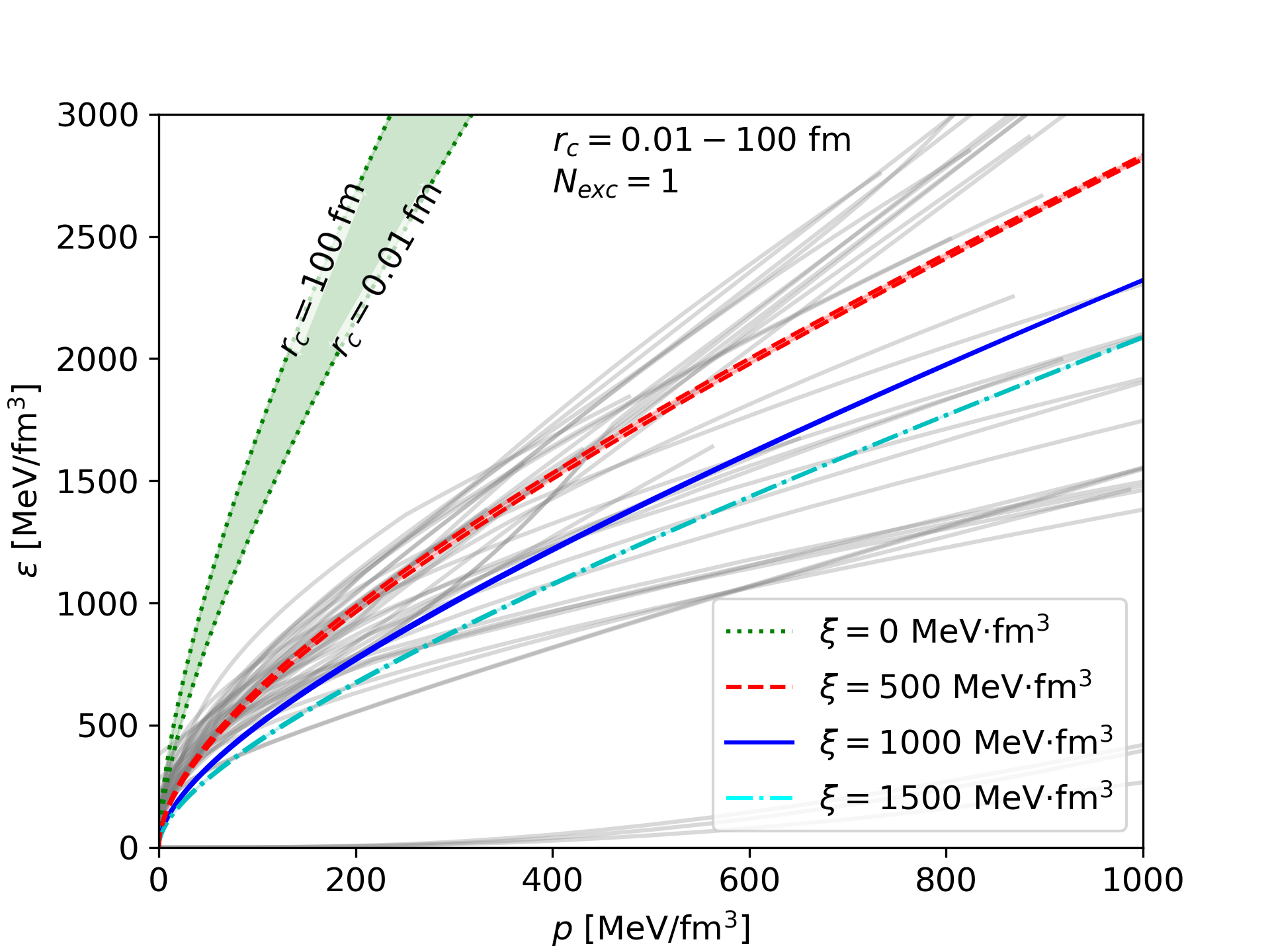}
        \label{fig:sky only} 
     \end{subfigure}
     \hfill
     \begin{subfigure}
         \centering
         \includegraphics[width=0.45\linewidth]{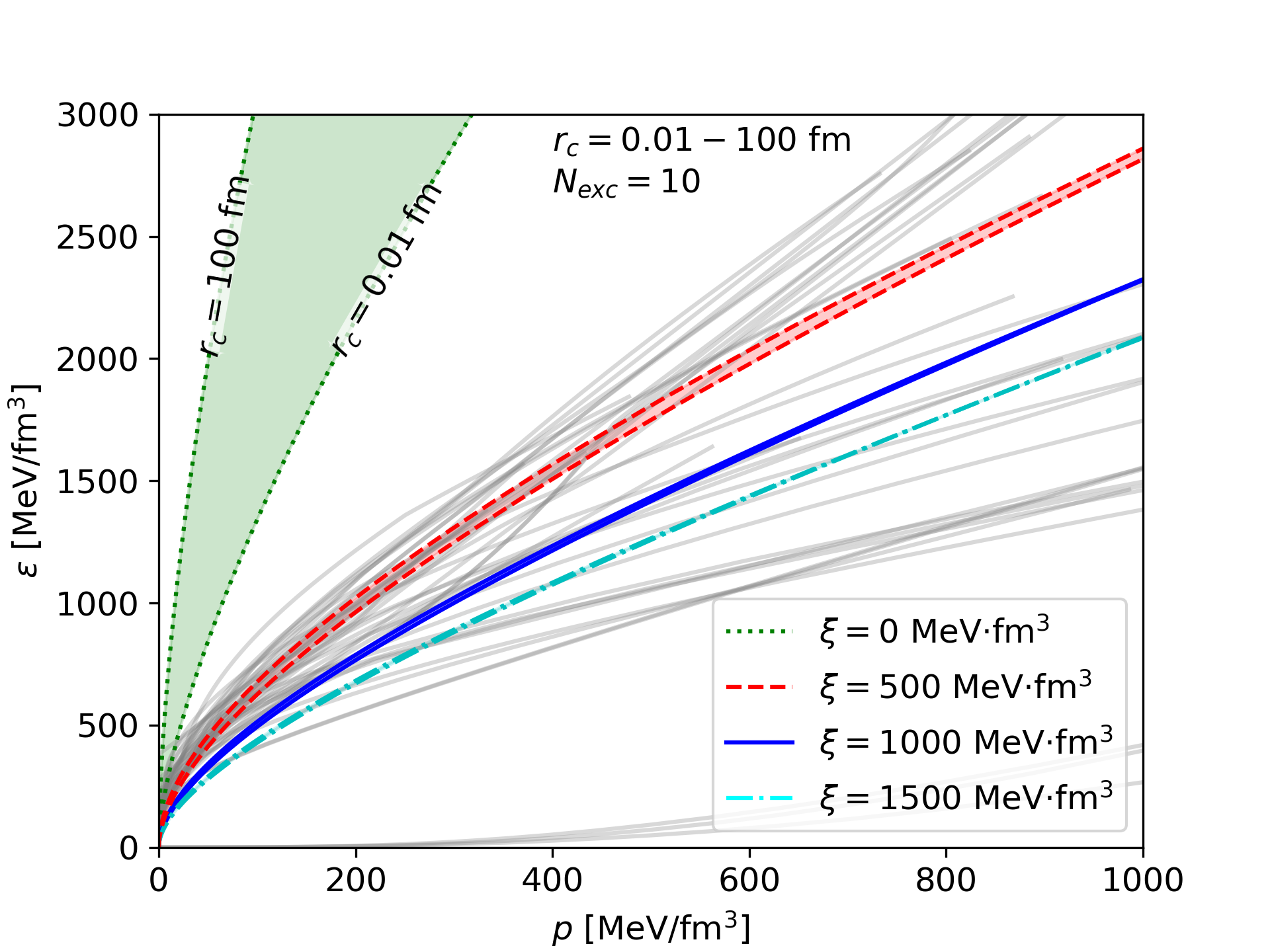}
         \label{fig:sky and water}
     \end{subfigure}     
     \hfill

     \hfill
     \begin{subfigure}
         \centering
         \includegraphics[width=0.45\linewidth]{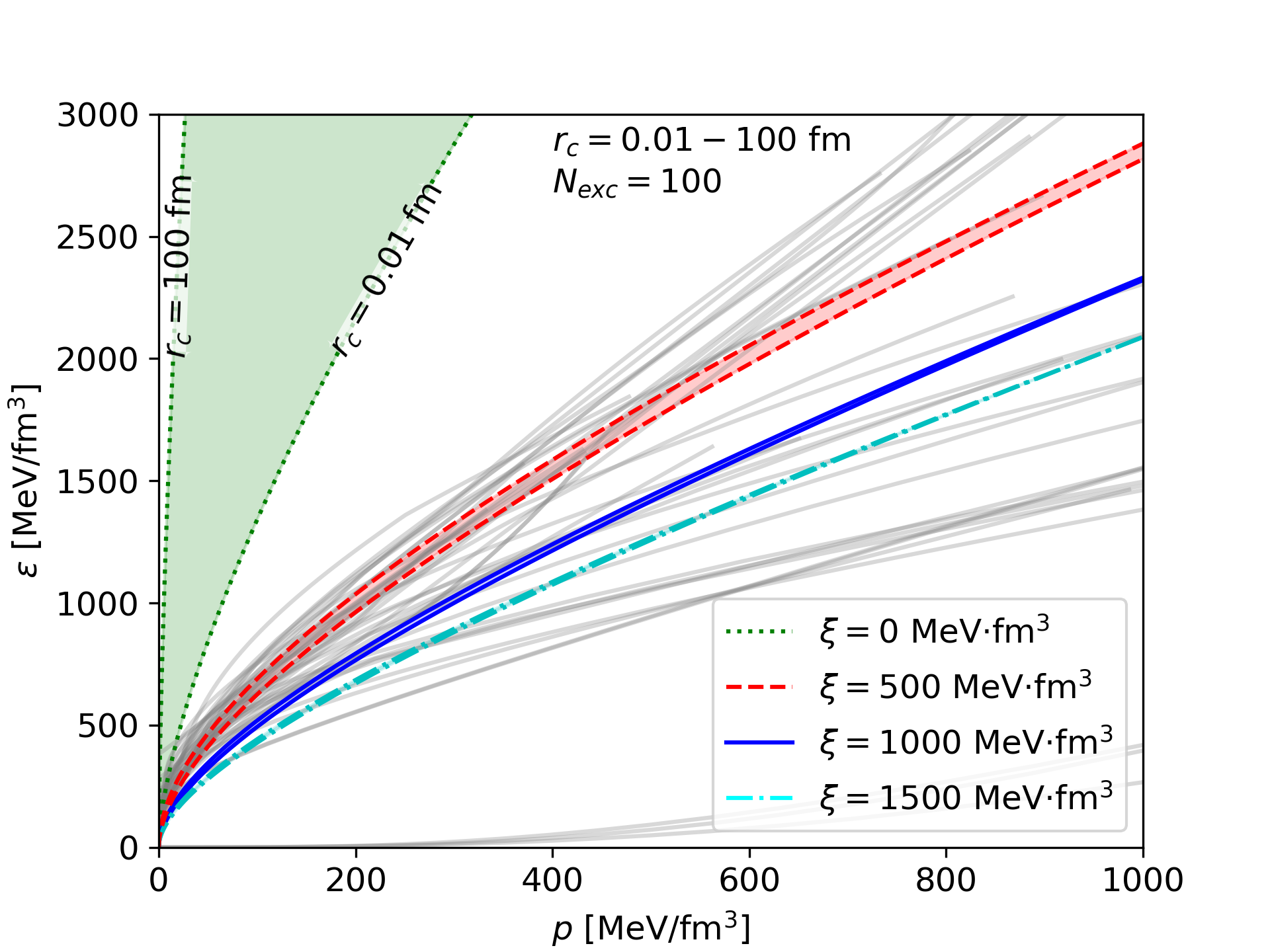}
     \end{subfigure}
     \hfill
     \begin{subfigure}
         \centering
         \includegraphics[width=0.45\linewidth]{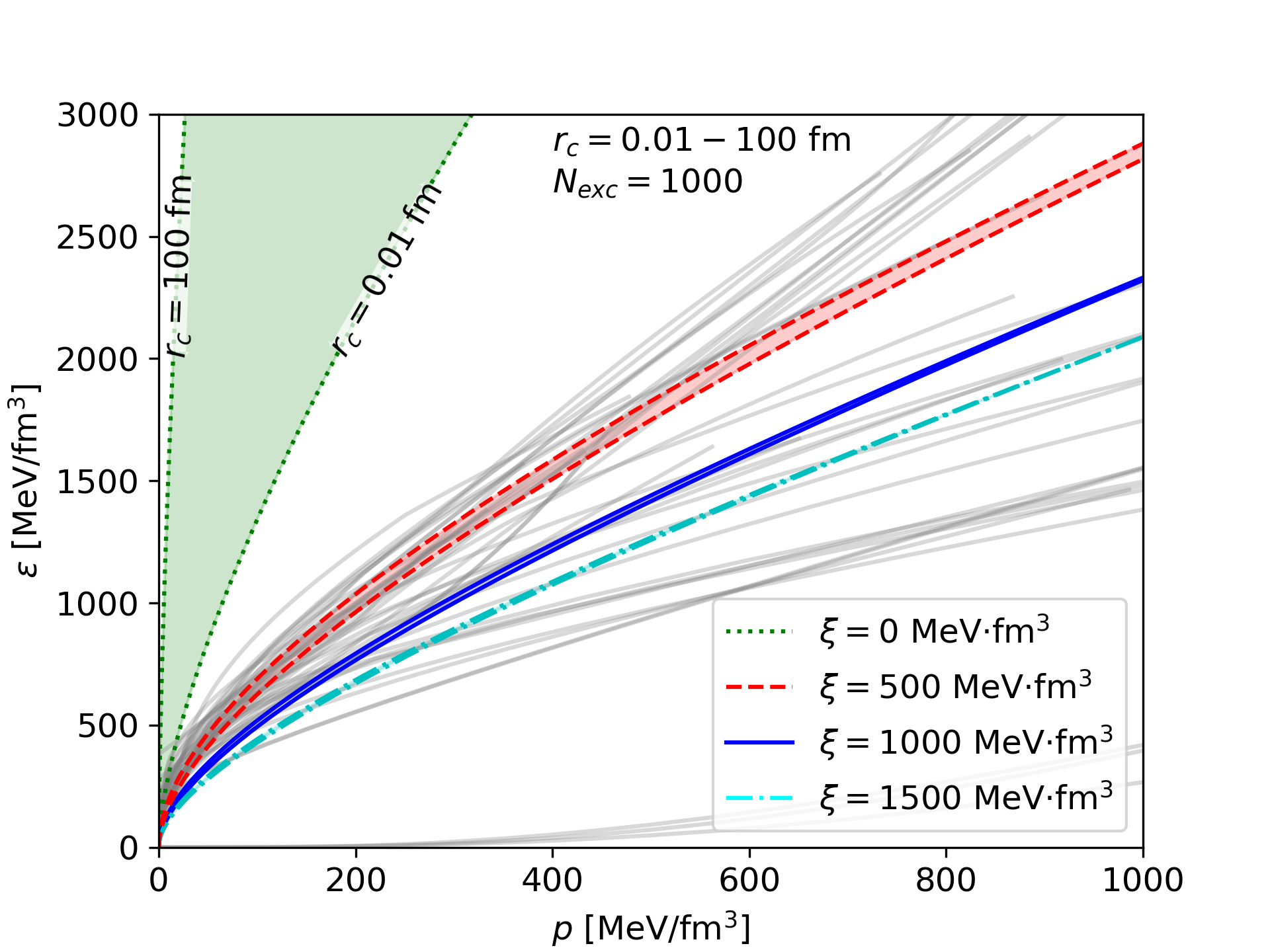}
         \label{fig:with coast}
     \end{subfigure}     
     \hfill

\caption{Equation of state: energy density as a function of pressure. Colors (line styles) represent different values of $\xi$, while each bundle spans $r_c$ values from 0.01~fm to 100~fm. Panels show EoSs for different excitation numbers, $N_\mathrm{exc}=$ 1, 10, 100 and 1000. We show EoSs taken from \citet{Ozel:2016oaf,compose2} for comparison (light grey curves).}
     \label{fig:vid}
\end{figure*}

Figure~\ref{fig:sat} shows pressure as a function of (total) baryon number density on a logarithmic scale spanning several orders of magnitude. Here, $N_{\mathrm{exc}}$ is fixed to 10, and the colored bands (with different linestyles respectively) corresponding to $\xi=0, 500, 1000$ and 1500~MeV$\cdot$fm$^3$ span $r_c$ values from 0.01 to 100~fm, increasing upwards. A brown, dashed vertical line shows the value of the nuclear saturation density ($n_{\mathrm{sat}}\approx 0.16$~fm$^{-3}$). In neutron stars, the density becomes larger than this value: it can reach multiple times the saturation density, see in~\cite{Ozel:2016oaf}. In this figure, we also plotted EoSs taken form other works  (\cite{2001ApJ...550..426L}, \cite{PhysRevD.79.124032}, \cite{Ozel:2016oaf}, \cite{PhysRevD.73.024021}, \cite{Alford_2005}, \cite{PhysRevC.58.1804}, \cite{1997A&A...328..274B}, \cite{BALBERG1997435}, \cite{PhysRevC.60.024605}, \cite{2013A&A...560A..48P}, \cite{1996ApJ...469..794E}, \cite{FRIEDMAN1981502}, \cite{1985ApJ...293..470G}, \cite{PhysRevC.60.025803}, \cite{MUTHER1987469}, \cite{MULLER1996508}, \cite{PhysRevLett.61.2518}, \cite{PhysRevD.52.661}, \cite{Douchin:2001sv}, \cite{PhysRevC.38.1010},\cite{RIKOVSKASTONE2007341}), which provide similar solutions to ours within the energy scale of interest.
Figure~\ref{fig:vid} shows the equation of state  on a linear scale, for four different $N_\mathrm{exc}$ values. We also show EoSs taken from~\cite{Ozel:2016oaf,compose2} in gray curves. First, let us concentrate on a single graph only, for example the top right one. The energy density, $\varepsilon$ is plotted as a function of pressure, $p$ for different $\xi$ and $r_c$ values. Here, $N_\mathrm{exc}$ is set to 10, which means that the highest energy level, that is allowed to be filled is the tenth\footnote{Due to the Pauli's exclusion principle, the same level is not allowed to be filled with the same fermionic states. That would mean a further fermions, which are propagating in the fifth dimension must have distinct, (higher) excited momenta.}. Bands with each color (line style) correspond to a certain coupling strength, $\xi$, and they span $r_c$ values between 0.01 and 100~fm. $r_c$, just like before, increases towards the top/left edge of the bands. Thus, one can see that the larger the $\xi$, the stiffer the EoS, indeed, changing the energy density results in a more prominent change in $p$. This is quite intuitive, since a larger $\xi$ means a stronger repulsive interaction, so the enhanced pressure increase is natural. On the other hand, changing $r_c$ has the reverse effect: the EoS is softer for a larger compactification radius. This is explained by the Kaluza\,--\,Klein ladder in Fig.~\ref{fig:kk_ladder}, where one can see that a larger radius means smaller gaps between energy levels. Thus, new degrees of freedom open up much easier, which leads to the coexistence of different "species" of particles at lower energies; a possibility of a higher number, thus energy density with moderate pressure. Another thing to notice is that the larger $\xi$ is, the less effect changing $r_c$ has, meaning the interaction will dominate the effect of the extra dimension. 

In this article, we generally set $N_{\mathrm{exc}}$ to be 10, however, it is worth considering what effect it has on our results. In Figure~\ref{fig:vid}, panels correspond to different maximal excitation numbers, $N_{\mathrm{exc}}$. Energy levels can be filled up to this theoretical maximum. One can see that the change affects the upper boundary of the $r_c$ bands only, thus it makes them "open up". This is because for $r_c=100$~fm, higher and higher energy levels are easily available. It is not so for $r_c=0.01$~fm, however, where even the first excitation has a huge energy, approximately 20~GeV, which cannot be associated with any known particle mass by~\cite{PhysRevD.110.030001}. For $r_c=100$~fm, the difference in $N_{exc}$ is significant for 1, 10 and 100, however, 1000 is practically the same as the previous. It does not cause any noticeable changes, at least for the energies considered here, due to the limited density of a compact star, as we will see later. The spectrum for this compactification radius is basically continuous. 

\subsection{Solving the TOV equation in Kaluza\,--\,Klein space-time}
\label{sec:tov}

The equation of state gives the properties of matter, while the Tolman\,--\,Oppenheimer\,--\,Volkoff equation describes the balance between gravity and pressure inside a spherically symmetric, static, time-independent object. This statement is valid in the Kaluza\,--\,Klein space as well. In~\cite{Lukacs:2003fh,Barnafoldi:2007} we presented, how the solution for the 5-dimensional case can be reduced to the 
usual, 4-dimensional TOV,
\begin{eqnarray}
\label{eq:tov}
\frac{\dd p(r)}{\dd r} &=& - \frac{GM(r) \varepsilon(r)}{r^2} \times  \nonumber \\
&\times &
\left[ 1+ \frac{p(r)}{\varepsilon(r)} \right]
\left[ 1+ \frac{4 \pi r^3 p(r)}{M(r)} \right]
\left[ 1 - \frac{GM(r)}{r} \right]^{-1} ,
\end{eqnarray}
where $r$ is the radius from the object's center, $G$ is Newton's gravitational constant, while $M(r)$ is the mass function
\begin{equation}
\label{eq:mass}
M(r)= \int\limits_{0}^{r} \dd r' 4\pi r'^2 \varepsilon(r') \ .
\end{equation}
The first term is just the Newtonian one describing hydrostatic equilibrium, while the last three multipliers are corrections due to general relativity. 
In our calculations, the Kaluza\,--\,Klein spacetime was used in a generalized Schwarzschild solution, where we set the extra $g_{55}$ in metric~\eqref{eq:diagmetric} to a constant scalar field. Thus an ordinary differential equation is obtained, that one can solve numerically by integrating it with respect to the star's radius $r$. The EoS then provides the connection between $\varepsilon(r)$ and $p(r)$ at each step. Initial boundary conditions are needed: we chose the central energy density of the object ($\varepsilon(r=0) = \varepsilon_c$) and the pressure at its surface (zero pressure condition, $p(r=R)=0$). Starting from the center, we build up the star layer by layer, until a given, realistic minimal pressure value is reached as it approaches zero. 
\begin{figure*}
    \centering
    \includegraphics[width=0.9\linewidth]{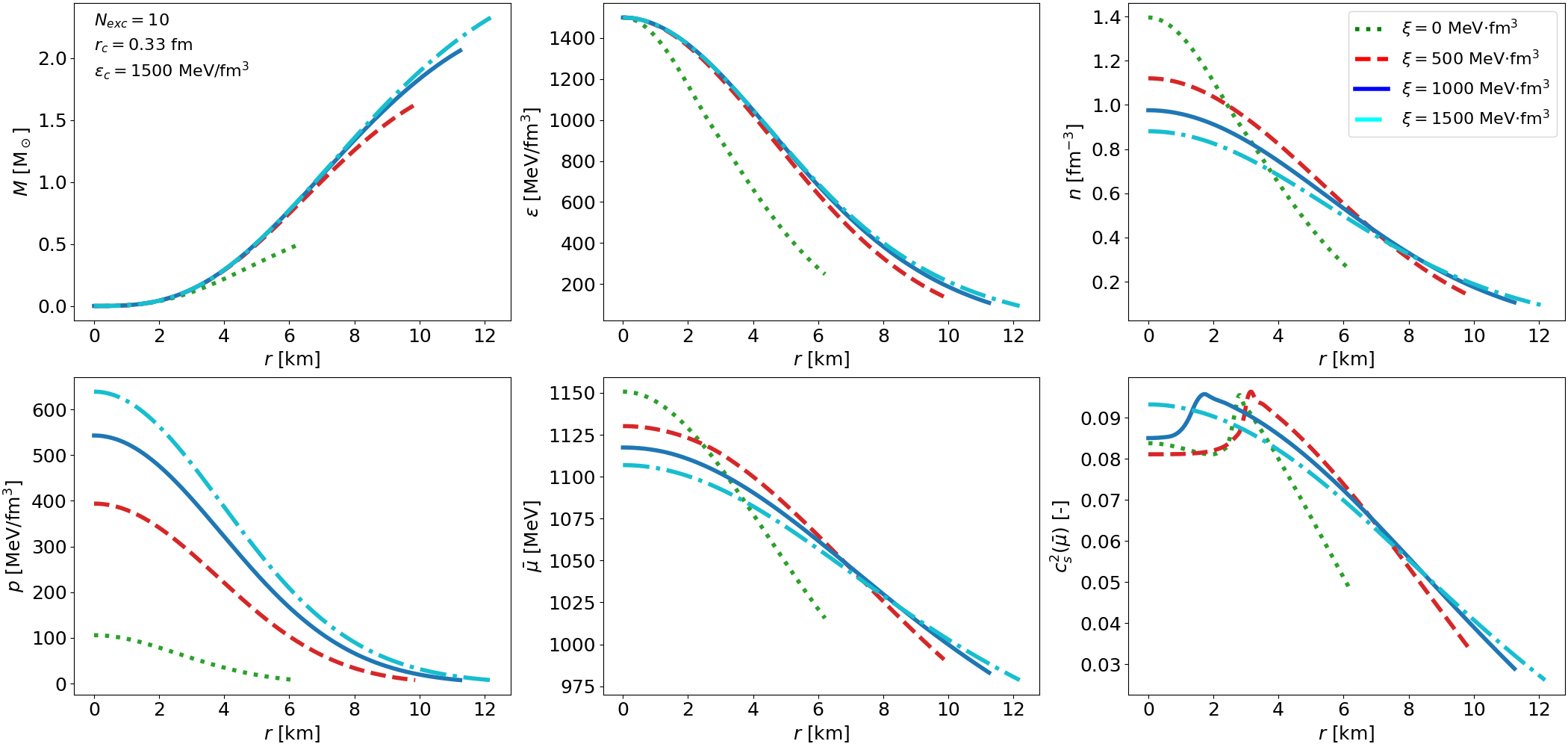}
    \caption{Mass and thermodynamical variables (energy density, number density, pressure, chemical potential and sound speed squared) as a function of $r$ for a compact object with central energy density $\varepsilon_c = 1500$~MeV/fm$^3$ and for fixed $r_\mathrm{c}=0.33$~fm and $N_{\mathrm{exc}}=10$. Colors and line styles correspond to different interaction coefficients, $\xi$ as above.}
    \label{fig:alljelr}
\end{figure*}

Figure~\ref{fig:alljelr} shows 
the structure of a star with central energy density $\varepsilon_c=1500$~MeV/fm$^3$ calculated with different EoSs -- $r_\mathrm{c}=0.33$~fm, $N_{\mathrm{exc}}=10$, while the interaction strength is changing: $\xi=$0, 500, 1000 and 1500~MeV$\cdot$fm$^3$. One can see that the harder the EoS, the larger the star -- both in mass and radius. Besides the mass, we show other quantities as a function of radius: energy density, $\varepsilon (r)$, baryon number density, $n(r)$, pressure, $p(r)$, chemical potential modified by the repulsive potential between particles, $\bar{\mu}(r)$ and the sound speed squared, $c_\mathrm{s}^2=\frac{n}{\bar{\mu}}\frac{\dd \bar{\mu}}{\dd n}$. Notice, that the curves for sound speed contain peaks -- as we move towards the center of the star, the quantity drops, as a new degree of freedom becomes available due to reaching larger energies. Furthermore, this switching is not visible in all cases, depending on the strength of the interaction, $\xi$, and also on the properties of the energy level in question. 

The numerical integration was performed by Python's {\tt scipy.integrate.solve\_ivp} using the RK45 explicit order 5(4) Runge\,--\,Kutta method.

\subsection{Boundary conditions at the surface of the compact object}
\label{sec:stopping}

The linear potential that we use, given by Eq.~\eqref{eq:poti} is most accurate at the largest densities, where the repulsive regime of the nuclear force dominates. This corresponds to the distance between hadrons being less than about $r_0\approx 0.8$~fm, where the minimum of the nuclear potential lies. Using an empirical approximation for the density, indeed one nucleon in a sphere with radius $r_0$,
\begin{equation}
    n \approx \left(\frac{4}{3} \pi r_0^3\right)^{-1} \approx 0.5~\text{fm}^{-3}
\end{equation}
and considering the nuclear saturation density $n_{\mathrm{sat}}\approx 0.16$~fm$^{-3}$, we set the surface of the star such that the number density at $R$ would be on the same order of magnitude as these. Although we use the zero pressure boundary condition, we chose to stop integrating the TOV equation at a slightly larger pressure value, $p(r=R)=10^{-5}$~km$^{-2}\approx7.56$~MeV/fm$^3$ to discard the effects of the EoS at the surface, based on the above arguments~\footnote{Where the pressure is described in units of km, $c$ and $G$ are set to 1.}. For example, with a compactification radius $r_c=0.33$~fm, an interaction strength, $\xi=1000$~MeV$\cdot$fm$^3$ and an excitation number, $N_{\mathrm{exc}}=10$, this pressure condition gives the number density at the surface as $n(r=R)\approx 0.11$~fm$^{-3}$. Using this condition, the maximal mass decreases with roughly 15\% for the same model parameters, however we argue that this gives a more accurate result, taking into consideration both the $M-R$ relation and the estimation of coefficients. Outer layers do not contribute much to the mass and radius of compact objects, since the EoS generally becomes softer at the surface, thus the density drops rapidly. 

\subsection{Kaluza\,--\,Klein star results and their comparison to observational and theoretical data}
\begin{figure}
    \centering
    \includegraphics[width=0.95\linewidth]{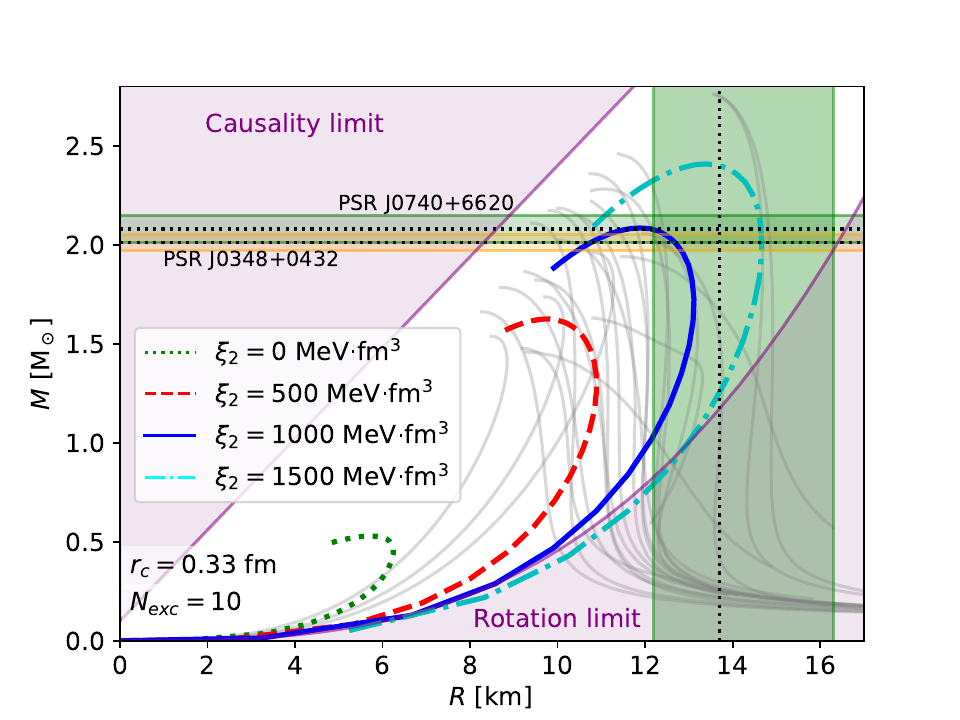}
    \caption{Mass\,--\,radius relations of compact stars using the model presented here \,--\, colors (linestyles) denote different $\xi$ values, while $r_\mathrm{c}=0.33$~fm and $N_{\mathrm{exc}}=10$. Measurement data for pulsars J0740+6620 (mass from \citet{Fonseca_2021} and radius from \citet{Miller:2021qha}) and J0348+0432 (mass from \citet{Antoniadis:2013pzd}) is shown in colored bands -- green and orange, respectively. The causality and rotation limits are shown in purple bands. We show the $M$--$R$ relatins of EoSs taken from \citet{Ozel:2016oaf,compose2} for comparison (light grey curves).}
    \label{fig:measure}
\end{figure}
In Figure~\ref{fig:measure} we present $M-R$ diagrams calculated using the method described above. Here the compactification radius of the extra dimension, $r_\mathrm{c}$ is fixed at 0.33~fm, while the excitation number, $N_{\mathrm{exc}}=10$. The interaction strength, $\xi$ is changing with values 0, 500, 1000 and 1500~MeV$\cdot$fm$^3$. The causality and rotation limits are shown in purple exclusion bands. We plotted the $M$--$R$ relations of EoSs taken from \cite{Ozel:2016oaf,compose2} for comparison (light grey curves). Measurement data for two pulsars, with about $2M_{\odot}$ mass, is shown : the mass from~\cite{Fonseca_2021} and the radius from~\cite{Miller:2021qha} for J0740+6620, and the mass from~\cite{Antoniadis:2013pzd} for J0348+0432. Although we show and examine the radius of stars calculated using our model, and compare it to measurement data, we stress that, according to Section~\ref{sec:stopping}, the mass is a more accurate observable for us. This is also true for observations, since measuring the radius of compact objects is a difficult task, and depends highly on the surface emission model used, see eg. in \cite{Miller:2016pom}. Nonetheless, one can see that with the right choice of interaction strength, $\xi$, we are able to reproduce both the radius and the mass of real objects. Here the two pulsars J0740+6620 and J0348+0432 were chosen to represent the maximum of the $M-R$ curve, as they are from the massive end of the currently known compact star population from Table~\ref{tab:maxmass}. Essentially, this is what we are interested in, since the upper limit for the compact star mass gives a strong constraint on the equation of state, especially in the higher density regime, see eg. in~\cite{Ozel:2016oaf} and~\cite{Lattimer:2019eez}.    


\section{Discussion}
\label{sec:disc}

Neither the size, nor the number of extra compactified dimensions is a directly observable quantity in a 
compact star. Since they exist only at a microscopical level of the theory, only quantum scale reactions would be able to make their effects directly measurable. However, this situation is standard in the investigation procedures of neutron star interiors, where the microscopical parameters of the cold dense nuclear matter can not be tested directly. The presented theoretical correspondence between the geometrical structure of the fifth dimension and known baryon states led us to check if deviation from known physical behavior may appear on a macroscopic scale -- as it was suggested and tested earlier in case of a non-interacting Fermi gas~\cite{Lukacs:2003fh,Barnafoldi:2007}.

The main measurable macroscopic parameters are the ones, which can provide constraints on the 5-dimensional general relativistic formalism, and on the nuclear matter structure via the equation of state. The physical interpretation and effects of the extra dimensional compactification radius and the maximal allowed number of excited states have already been investigated in~\cite{Barnafoldi:2015wca,Karsai:2016wfx}.
We have found here that by inserting a simplistic one-parameter interaction, the above-mentioned correspondence between the geometrical structure of the extra dimension and baryon states is still valid. Furthermore, since the obtained compact star observables become realistic, this provides us the opportunity to constrain the model in a phenomenological way. 

The maximal mass ($M_{\textrm{max}}$) of compact stars is one of the observables, which can constrain the parameters of the nuclear matter (EoS) very well, as pointed out in~\cite{Posfay:2020xgp, Barnafoldi:2020phq} first, and applied to compressibility estimates by Bayesian methods in the $\sigma-\omega$ model in~\cite{Alvarez-Castillo:2020fyn,Alvarez-Castillo:2020aku,Szigeti:2021opd} later. Here, we follow the same strategy, and focus only on the Kaluza\,--\,Klein compact star's maximal mass with respect to the spacetime structure of the model and the parameters of the interacting Fermi gas. 
\begin{figure}
    \centering
    \includegraphics[width=0.95\linewidth]{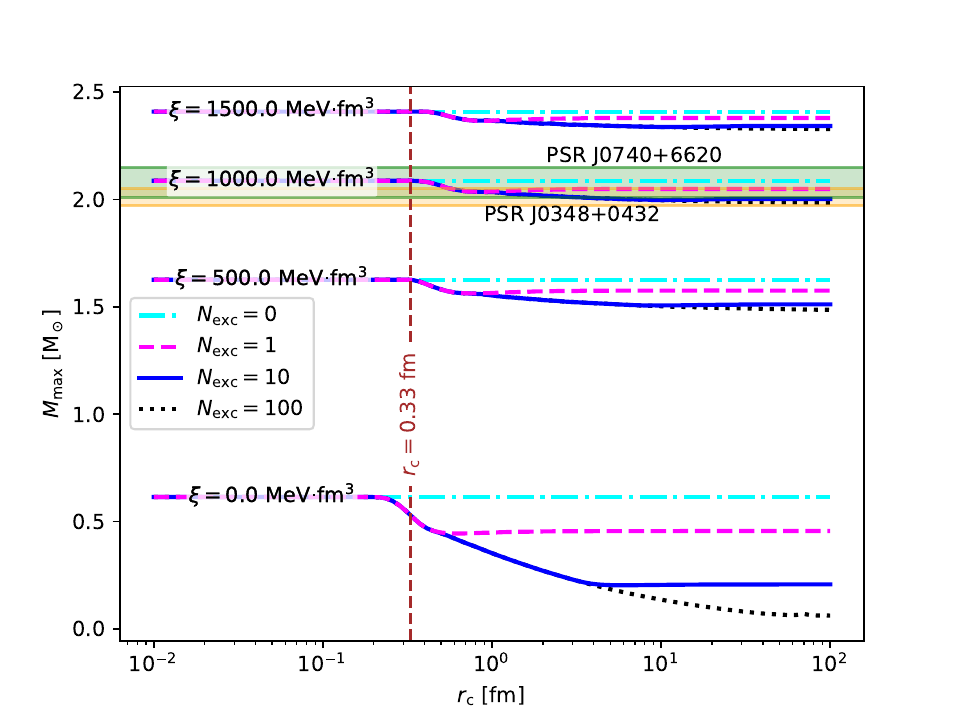}
    \caption{Maximal mass of compact stars as a function of $r_\mathbf{c}$ shown for different $\xi$ values and excitation numbers (indicated by color and linestyle). Colored bands show mass measurement data for pulsars J0740+6620 from \citet{Fonseca_2021} and J0348+0432 from \citet{Antoniadis:2013pzd}. The $r_{\mathrm{c}}= 0.33$~fm value is shown as reference by a brown dashed line.}
    \label{fig:bird}
\end{figure}

The independent parameters of our model are: the strength of the repulsive interaction, $\xi$, the size of the extra compactified dimension, $r_\mathrm{c}$ and the maximum allowed excitation number, $N_\mathrm{exc}$. Each one of them has an effect on the maximal mass of stars. However, their contribution hugely differs, indeed their interplay is not straightforward. In Figure~\ref{fig:bird}, we plotted the maximal mass of stars corresponding to different sets of parameters: $\xi=$~0, 500, 1000 and 1500~MeV$\cdot$fm$^3$, $N_\mathrm{exc}=$~0, 1, 10 and 100 (denoted by colors and linestyles), while the  horizontal axis corresponds to $r_\mathrm{c}$ running from 0.01 to 100~fm on a logarithmic scale with 100 discrete values for calculations. Colored bands show mass measurement data for pulsars J0740+6620 from~\cite{Fonseca_2021} and J0348+0432 from~\cite{Antoniadis:2013pzd}. We note that the case $N_\mathrm{exc}=1000$ in this $r_\mathrm{c}$ regime coincides with the results of $N_\mathrm{exc}=100$ within numerical errors, thus here we omitted plotting it. The maximal mass for each parameter set is determined by calculating the $M-R$ curve consisting of 50 stars, fitting a spline function to it, then taking its maximum where the derivative vanishes.  In the following, we examine the effects of each parameter:

\begin{description}
    \item[The interaction strength, $\xi$] has the largest effect on the maximal mass of compact stars -- it basically produces a vertical shift and compression of the curve bundles, whose structure is formed by the other two parameters. We notice, this property is inherited from the scaling property of the Tolmann\,--\,Oppenheimer\,--\,Volkoff equations, via the shift in the pressure term~\cite{Glendenning:1997wn}, manifest in the magnification of both observables on the $M$-$R$ diagram. As in Section~\ref{sec:eos}, here in Figure~\ref{fig:bird} as well it becomes evident that the larger $\xi$ is, the less effect $r_\mathrm{c}$ and $N_\mathrm{exc}$ have. The $\xi$ parameter dominating the resulting observables is clear, since its effects appear throughout the star, while the other two parameters become important only at larger densities. Furthermore, nuclear forces, which the repulsive potential is meant to describe, play a significant role in the shaping of compact objects.   
    \item[The compactification radius, $r_\mathrm{c}$] is the second most important parameter. Its effects are dominant at the same values, regardless of the interaction strength, $\xi$. This is because a certain $r_\mathrm{c}$ size corresponds to a certain energy, as it is connected to the allowed momentum of particles in the extra dimension, and sets the magnitude of their effective mass, $\bar{m}$ via Eq.~\eqref{eq:mass}. Energies that are relevant for neutron stars thus have a more significant effect. At very small compactification radius values, $r_\mathrm{c} \lesssim 0.2$~fm, the energy levels corresponding to the first excitation, $N_\mathrm{exc}=1$ are so large, that they basically do not appear in compact stars, at least not in a notable amount. For $r_\mathrm{c} \gtrsim 1$~fm, the evolution with respect to $r_\mathrm{c}$ is quite steady -- the maximal mass slowly decreases. For large $r_\mathrm{c}$ values the energy spectrum essentially becomes a continuum, thus the appearance of a new degree of freedom does not cause a qualitative difference.  
    \item[The maximal possible excitation number, $N_\mathrm{exc}$] is less physical than the other two parameters. Its value could be taken to be infinity, practically $N_\mathrm{exc} \approx 100$, however, for most of our investigations $N_\mathrm{exc} = 10$ was enough, since the relevant energy for compact stars, where $r_\mathrm{c}$ is between about 0.2~fm and 2.0~fm does not produce excitations higher than $N_\mathrm{exc} = 10$. This is shown by the fact that the curves corresponding to $N_\mathrm{exc} = 10$ and $N_\mathrm{exc} = 100$ in Fig.~\ref{fig:bird} coincide within this $r_\mathrm{c}$ regime, and separate only for larger compactification radii. However, when a new energy level would become available inside a star, but $N_\mathrm{exc}$ is fixed at a lower value, the maximal mass curves become constant as $r_\mathrm{c}$ increases. For the case when $N_\mathrm{exc} = 0$, $r_\mathbf{c}$ has no effect, naturally.
\end{description}  

Finally we can compare our model parameters to neutron star observables, taking into account that by our recent knowlegde from~\cite{Breu_2016,musolino2023maximummassoblatenessrotating}, pulsars J0740+6620~\cite{Fonseca_2021} and J0348+0432~\cite{Antoniadis:2013pzd} likely exhibit close-to-maximal mass. With this assumption, in Figure~\ref{fig:bird} we can read off the correct interaction strength, $\xi \approx 1000 \pm 200$~MeV$\cdot$fm\textsuperscript{3} corresponding to our Kaluza\,--\,Klein model with interacting Fermi gas. As a consequence of 
our phenomenological inputs, effects of the compactification radius are visible only if $r_\mathrm{c} \gtrsim 0.2$~fm, and the momentum in the direction of the compactified extra dimension is able to fill several excited states, $N_\mathrm{exc} \ge 1$. Taking into account the maximal mass measurement uncertainty of about 5\%, one can see that the precision of the $\xi$-estimate does not allow the further restriction of the size of the extra compactified dimension, or the number of possible excited states. However, by fixing $\xi$ based on nuclear physics arguments, and with the accumulation of more precise observational data, a possibility to restrict $r_\mathrm{c}$ and $N_\mathrm{exc}$ opens. 
\begin{figure}
    \centering
    \includegraphics[width=0.95\linewidth]{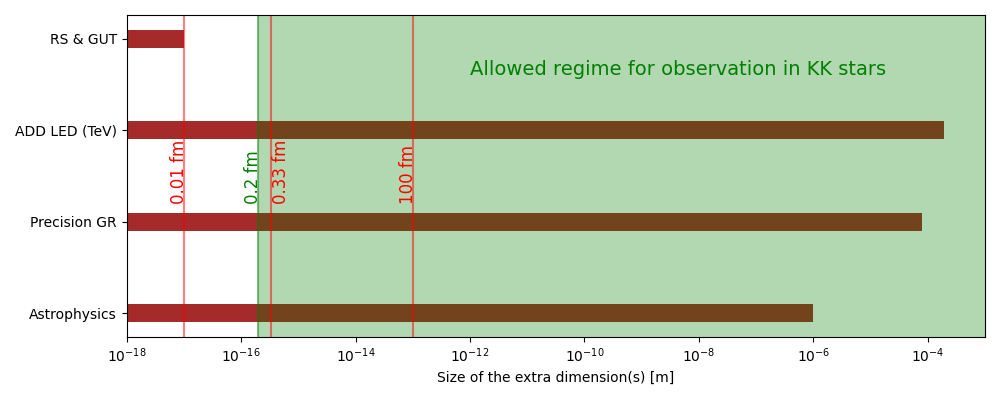}
    \caption{Allowed regime for observation in the Kaluza\,--\,Klein compact star,  $r_\mathrm{c} \gtrsim 0.2$~fm (green color). Brown bars show the model-predicted values of the size of the extra compacified dimensions' measurement data. The $r_{\mathrm{c}}= 0.01, 0.33$ and $100$~fm values are shown as reference by a red line.}
    \label{fig:excl}
\end{figure}


\section{Summary and Conclusions}
\label{sec:sum}

Compact stars within a Kaluza\,--\,Klein-based theory, where spacetime is extended by the addition of one extra compactified spatial dimension have been investigated. Due to the smallness of the added dimension -- on the order of a few fm -- particles need to reach a certain energy to be able to propagate in it, with momenta quantized by the finite size volume of the extra 5\textsuperscript{th} direction. In order to get the main astrophysical observables of compact stars -- their mass and their radius --, the numerical integration of the Tolman\,--\,Oppenheimer\,--\,Volkoff equation was performed, with an interacting Fermi gas providing the equation of state. A key element of the current work was to show that by introducing a simple, linear, repulsive potential to the model already discussed in~\cite{Barnafoldi:2015wca} and~\cite{Karsai:2016wfx}, compact stars with realistic mass--radius relations can be simulated. Once the other two model parameters, the interaction strength, $\xi$ and the maximal excitation number, $N_\mathrm{exc}$ are fixed based on empirical and theoretical arguments, constraints on the existence and size of extra compactified spatial dimensions could be given by comparison to measurement data of close-to-maximal pulsars J0740+6620~\cite{Fonseca_2021} and J0348+0432~\cite{Antoniadis:2013pzd}. 

Finally, we have found that the observationally constrained interaction strength, $\xi \approx 1000 \pm 200$~MeV$\cdot$fm\textsuperscript{3} does not only provide a realistic maximal mass star, but its value is in agreement with a similar cold and dense nuclear matter model presented in~\cite{Zimanyi:1987bt} at the order level. 
Furthermore, we have shown that it is indeed possible to constrain the model parameters -- such as the radius, $r_\mathrm{c}$ -- of extra dimensional theories based on neutron star measurements. In our case, the observational regime is where $r_\mathrm{c} \gtrsim 0.2$~fm, which was found to be compatible with other theoretical model predictions on extra dimensions as they are compared in Figure~\ref{fig:excl}.


\section*{Acknowledgements}

Authors gratefully acknowledges the Hungarian National Research, Development and Innovation Office (NKFIH) under Contracts No. OTKA K135515 and K147131, No. NKFIH NEMZ\_KI-2022-00031 and Wigner Scientific Computing Laboratory (WSCLAB, the former Wigner GPU Laboratory). Author A. H. is supported by NKFIH through the DKÖP program of the Doctoral School of Physics of Eötvös Loránd University and HUN-REN's Mobility fellowship with indentifiers KMP-2023/101 and KMP-2024/31. E.F-D. also received funding from the NKFIH excellence grant TKP2021-NKTA-64. The authors acknowledge the fruitful discussions with Balázs Asztalos, Szilvia Karsai, Péter Lévai, and Aneta Magdalena Wojnar. 

\section*{Data Availability}

The data underlying this paper will be shared on reasonable request to the authors.




\bibliographystyle{mnras}
\bibliography{KKstar} 




%
%
%

\bsp	
\label{lastpage}
\end{document}